\documentclass[manuscript,screen,natbib=false]{acmart}
\usepackage{multirow}

\usepackage[style=acmnumeric]{biblatex}
\AtBeginDocument{%
  \providecommand\BibTeX{{%
    \normalfont B\kern-0.5em{\scshape i\kern-0.25em b}\kern-0.8em\TeX}}}

\copyrightyear{2022}
\acmYear{2022}

\begin{document}

\title{A Review on Pushing the Limits of Baseline Recommendation Systems with the integration of Opinion Mining \& Information Retrieval Techniques}

\author{Dinuka Ravijaya Piyadigama}
\email{drpiyadigama@gmail.com}
\orcid{0000-0002-8123-0099}
\affiliation{%
  \institution{University of Westminster}
  \streetaddress{309 Regent St.}
  \city{London}
  \country{UK}
}

\author{Guhanathan Poravi}
\affiliation{%
  \institution{Informatics Institute of Technology}
  \streetaddress{57 Ramakrishna Rd}
  \city{Colombo 06}
  \country{Sri Lanka}}
\email{guhanathan.p@iit.ac.lk}


\begin{abstract}
Recommendations Systems allow users to identify trending items among a community while being timely and relevant to the user's expectations. When the purpose of various Recommendation Systems differs, the required type of recommendations also differs for each use case. While one Recommendation System may focus on recommending popular items, another may focus on recommending items that are comparable to the user's interests. Content-based filtering, user-to-user \& item-to-item Collaborative filtering, and more recently; Deep Learning methods have been brought forward by the researchers to achieve better quality recommendations. 

Even though each of these methods has proven to perform well individually, there have been attempts to push the boundaries of their limitations. Following a wide range of methods, researchers have tried to expand on the capabilities of standard recommendation systems to provide the most effective recommendations to users while being more profitable from a business's perspective. This has been achieved by taking a hybrid approach when building models and architectures for Recommendation Systems.

This paper is a review of the novel models \& architectures of hybrid Recommendation Systems. The author identifies possibilities of expanding the capabilities of baseline models \& the advantages and drawbacks of each model with selected use cases in this review.
\end{abstract}

\ccsdesc[500]{Information systems~Recommender systems}
\ccsdesc[500]{Human-centered computing~Social recommendation}
\ccsdesc[300]{Information systems~Retrieval models and ranking}
\ccsdesc[300]{Information systems~Data mining}
\ccsdesc[100]{Applied computing~Online shopping}

\keywords{Recommendation Systems, Collaborative Filtering, Content based Filtering, Hybrid Recommendation Systems, Opinion Mining}

\maketitle

\section{Introduction}
In the modern-day age, Recommendation Systems play a vital role in almost every B2C and B2B system. These systems aid in the resolution of the problem of information overload.

Recommending items for purchase, displaying personalized recommendations for users to watch videos/ movies, displaying advertisements to users, displaying personalized recommendations for online profiles and content on social networks, displaying the most likely-to-use tools/ software in a system are all done using Recommendation Systems. In 2018 it was estimated that 35\% of Amazon's revenue \cite{naumov_deep_2019} is driven by Recommendation Systems. 75\% of Netlfix viewer activity \cite{vanderbilt_science_nodate} was also said to come from recommendations back in 2013. There are many types of standard recommendation algorithms \& systems made to cater to these use cases. 

Among the many types of recommendation systems, item-to-item Collaborative filtering has been the most successful technique, while user-to-user Collaborative filtering and Content-based filtering have also had their own upsides. In order to take advantage of the relevant advantages of each method, Hybrid recommendation systems were introduced. More recently; Deep Learning methods have been brought forward by researchers to achieve better quality recommendations.

Even though each of these methods has proven to perform well, there have been attempts to push the boundaries of their limitations. Following a wide range of methods, researchers have tried to expand on the capabilities of standard recommendation systems in order to provide the most effective recommendations to users while being more profitable from a business's perspective. This has been achieved by ensembling models or taking hybrid approaches when building models and architectures for Recommendation Systems.

\section{Machine Learning-based recommendation techniques}
There are several baseline techniques of Recommendations Systems that have been used by the biggest data-driven companies around the world.
Among the many types of recommendation systems, \textbf{item-to-item Collaborative filtering} \cite{linden_amazoncom_2003} has been the most successful technique for an extended period of time \cite{smith_two_2017}, while user-to-user Collaborative filtering and Content-based filtering have also had their own upsides. In order to take advantage of the relevant advantages of each method, Hybrid recommendation systems \cite{geetha_hybrid_2018} were introduced.

\section{Deep Learning-based recommendation techniques}
In 2019, \textbf{Facebook} open-sourced a new categorical data-driven \textbf{Deep learning-based recommendation engine} \cite{naumov_deep_2019, noauthor_we_2019}. This recommendation model was developed from the two perspectives of recommendation systems and predictive analytics. It made use of embeddings, two Multilayer Perceptrons (MLPs), one sigmoid function, \cite{freudenthaler_factorization_2011} and a parallelization scheme to support large scales of data.

More recently, after many attempts to go beyond the gold standard of recommendation systems \cite{linden_amazoncom_2003, smith_two_2017} with the use of deep learning techniques, Amazon finally has achieved to use an \emph{"Auto-Encoder"} Deep Neural Network to give better movie recommendations \cite{larry_history_2019}.

\section{Concerns about progress in Recommendation Systems}
In several research \& review papers, it has been brought to sight that Deep learning techniques in the area of recommendation systems have failed to live up to the expectations compared to the advancements in Computer Vision, Speech Recognition \& Natural Language Processing (NLP) domains \cite{choi_local_2021}. The results that have been published presenting advancements in the Recommendation Systems domain using Deep learning techniques have not been very convincing for the majority of use cases. Many standard Machine learning \& regression techniques have been able to outperform systems created using Deep learning models in terms of recommendations. As highlighted in past reviews \cite{dacrema_are_2019} it is understood that Deep learning models have been used as baseline methods for evaluating new Deep learning models. Thus, when looking back at older Machine learning techniques, they haven't been making an impactful improvement in many cases. As a result, much of the work related to Recommendation Systems using Deep learning techniques has been giving poorer recommendations, for higher computational power.

A study conducted in 2019 questioned if we are really making any progress with Deep Learning models in the domain of Recommendations \cite{dacrema_are_2019}. In a more recent study, researchers tried to understand the similarities and advantages of using \textbf{MLP} versus \textbf{dot product} \cite{rendle_neural_2020}. Similar to many Deep learning approaches, it was understood that MLP wasn't necessary unless the dataset was too large or the embedding dimension was very small. A dot product was identified as a better choice since it was efficient to a satisfactory extent.

\section{How to choose the ideal algorithm for a Recommendations System?}

Generally, an application of a Recommendation System will come in a business use case, where companies focus on maximizing profits for minimum expenses. In a scenario like that, it would make more sense to choose a cheaper model that gets the job done to a satisfactory level. Dot products offer a significant advantage over MLPs in terms of inference cost due to the availability of efficient maximum inner product search algorithms. Since MLPs are too costly to use in production environments, the better default choice in most cases would be the dot product approach that uses Machine Learning techniques with Matrix Factorization.

\begin{equation}
<x,y>  = \sum_{i=1}^{d}x_{i}y_{i}\label{eq:dot-product}
\end{equation}

\begin{equation}
y = \varphi(\sum_{i=1}^{n}w_{i}x_{i}+b) = \varphi(w^{T}x+b)\label{eq:single-layer-perceptron}
\end{equation}

The equation \ref{eq:dot-product} is used for the calculation of the dot product between two items' similarities, while the equation \ref{eq:single-layer-perceptron} is used in a single-layer perceptron, where w denotes the vector of weights, x is the vector of inputs, b is the bias and $\varphi$ is the non-linear activation function.

A variation that combines the MLP with a weighted dot product model, named \textbf{\emph{Neural Matrix Factorization (NeuMF)}} has also been explored. But, that too is deemed to be outperformed by the dot product method.

One of the major limitations identified related to dot products in this study is that learning a dot product with high accuracy for a large embedding dimension required a large model capacity. This may also require more computational resources. Therefore, it would be advisable for Data Science engineers to consider both approaches based on the requirements \& data of the system that they're planning to work on.


\section{User Opinion \& Sentiment Aware Recommendation Systems}

\begin{quote} 
\centering 
"Users usually transmit their decisions together with emotions."\cite{chen_user_2019}
\end{quote}

User emotions are an important factor to be considered when trying to get a better understanding of the probable decisions of a user. Sentiment analysis of user-generated content can be ideal to provide users with better recommendations. Opinion mining is a process to identify another person's viewpoint on something. Sentiment analysis is to extract someone's attitude or feeling \cite{nah_opinion_2018}. Both these measures can be considered to be important in understanding a user's opinion related to an item.

One of the easiest ways to capture the opinions of users is to use content generated on social media. It's a cheap, fast, and effective way of capturing user opinions.


\subsection{Extracting User Review Sentiment for Recommendations}
There have been studies to understand the influence of user sentiment on the use of user reviews.
Sentiment Analysis techniques have been applied for the purpose of understanding users' opinions related to movies that they have previously watched, in order to understand the user's preference profile. In previous research, \cite{cheng_hybrid_2020} a framework that is capable of summarizing an aggregated list of historical reviews given by a user has been introduced. Later, these results are combined together with a Collaborative Filtering algorithm. This overcomes the problem of 'data sparsity' which occurs as a result of depending on user ratings. 
Another advantage that this method provides is that it enables the system to help creators identify the preferences of the movie consumers.

One of the major concerns that this method seems to have is that while it will be able to give appropriate recommendations to a user, the recommendations will most likely contain old movies that have been watched by other users as well. The system will have a difficulty in identifying new movies that are highly trending.

The strategy of the framework that has been adapted here has not considered different aspects of reviews that users may place. A mentioned example is \emph{the user might focus on the quality of the sound effects in action movies, but the storyline in dramas}. The semantic strategy of opinion extraction is noted as an area to be worked on. Another limitation that has been mentioned in relation to this framework is that it does not consider slang, irony, or sarcasm, although these styles of semantics have the ability to completely change the person's opinion.

Because the proposed method is highly reliant on the technique of text-mining user reviews, the final recommendations could have a positive influence if greater attention is paid to research on text-mining models and relevant procedures.

\subsection{Cross-Domain Recommendations with Decision-making Support based on Twitter Sentiment}
\subsubsection{A Cross-Domain Hybrid Recommendation System}
When it comes to recommending items from multiple domains, it becomes challenging to use the same recommendation model for recommendations. Each of these domains may have distinctive features as well as a varied bias in weights of each feature towards recommendations.
Taking into consideration of the 3 domains music, movies \& books a group of researchers has been able to produce a cross-domain recommendations system \cite{ayushi_cross-domain_2018}. This work has focused on using the domain knowledge gained from movies to generate recommendations for books and music.

In order to build the required system, the authors have tested various supervised classification algorithms together with a hybrid approach for recommendations with the combination of content-based recommendations, user-to-user Collaborative filtering, and personalized recommendation techniques. Out of the tested classifiers, the Decision tree classifier was found to give the highest accuracy.

The system that was produced in this research was able to address the limitation of data sparsity and the cold start problem that occurs with single domain analysis. With the integration of several domains, the system has shown the capability of generating a higher accuracy in suggestions.

\subsubsection{Using Twitter Sentiment for Validation of Recommendations}
The authors have then taken the extra step of using Twitter sentiment analysis on the generated recommended entities. One of the points that can be taken out of this is that public sentiment on social media is a consideration that users show interest in when choosing an item to consume, even after getting it as a recommendation. This drills down to the natural human desire to get validation on consumables by the people around them and from those that they look up to. Taking this into consideration for a recommendation system is, therefore, a positive aspect, especially for systems that don't have the luxury of integrating directly with a large social network such as Facebook/ Twitter in order to generate direct recommendations for its users.

The overall system uses this as a decision-support system to provide the user with decision making by visualizing the positive, negative \& neutral polarity percentages given by people, on Twitter. While it is clear that such decision support is valuable to the user, it feels ironic to recommend an item to a user and then say that it's not well-accepted by the community. While helping the user identify the recommended items based on other users' online activity, it may also say that it's not recommended, or deemed with negative sentiment by the majority of the community. This may leave the user in the confusion about what to take. It also leaves the part of the work the system can do for the user. A better method to integrate this public sentiment into recommendations would be to utilize the sentiment scores on social media to harvest the ideal recommendations and then show them to the user. Items with only positive sentiment can be considered.

\subsection{Identifying Possible Classification Techniques of User Reviews}

While procedures relevant to text-mining play an important role in understanding users' opinions, the classification of these reviews also has to be done appropriately, in order to be used for recommendations.

The emotional information that users provide with their comments has the potential to influence the correctness and precision of recommendations. In the work done \cite{chen_user_2019}, a deep learning model has been used to process user comments to generate a possible rating for recommendations.

Sentiment analysis is applied to the reviews to create a feature vector. Then, a noise reduction procedure is implemented on the data set to delete short comments, comments with no expression, and false rating comments. This has been done to improve the classification of ratings, as it has been pointed out in previous literature\cite{cheng_hybrid_2020} as being highly reliant on generating recommendations based on sentiment analysis of user-generated content. Finally, a Deep Belief Network has been used to achieve data learning for the recommendations.

This \textbf{Deep Belief Network and Sentiment Analysis (DBNSA)} has been said to outperform baseline models, especially training loss value, precision, and recall on Yelp and Amazon data sets. Furthermore, it is said to save more time than other baseline methods. The biggest drawback that this method seems to have is that the algorithm is not suitable for real-time testing.
Furthermore, social relationships and subsequent timeline comments have also been identified as a possible extension of this work since they can help address the cold-start problem by using timeline comments from a user's close social relationships. This might be a little difficult to implement by integrating externally into social networks since a user's close group of friends who think alike will have to be known to get the ideal information that will affect a particular user's decision-making process.

\bigbreak

There has been previous research focused on devising a robust recommendation methodology by identifying the credibility of reviewers and the quality of reviews when taking into consideration of all item reviews \cite{hu_reviewer_2020}. Sentiment analysis captured from reviews is an additional enhancement that this model uses apart from identifying the factors that affect a user's fondness for a certain product. This has been identified as the first work that integrates credibility-driven feature-based fine-grained sentiment analysis with user modeling for online product recommendations.

The entire system which is named credibility, interests, and sentiment enhanced recommendation (CISER) has five sub-modules. Candidate feature extraction is done using the \emph{spaCy} library, while sentiment confidence is given by \emph{fastText}. Thereafter, reviewer credibility analysis, user interest mining, candidate feature sentiment assignment, and recommendation module follow.
When the proposed system was tested with Amazon's camera review data-set, it managed to outperform baseline models. This shows that the more specifics we look into when considering items for recommendations, the better the Recommendations System is able to perform. 

The paper suggests devising sophisticated measures for representative rating and expertise as possible future enhancements that can be done to increase the specifics of identifying reviewer credibility. Furthermore, social network information and online activity logs can be considered for the user credibility model.




\section{Breakdown of Recommendation System Architectures that integrate opinion mining techniques}
\begin{figure}[h]
\centering
\includegraphics[width=\textwidth]{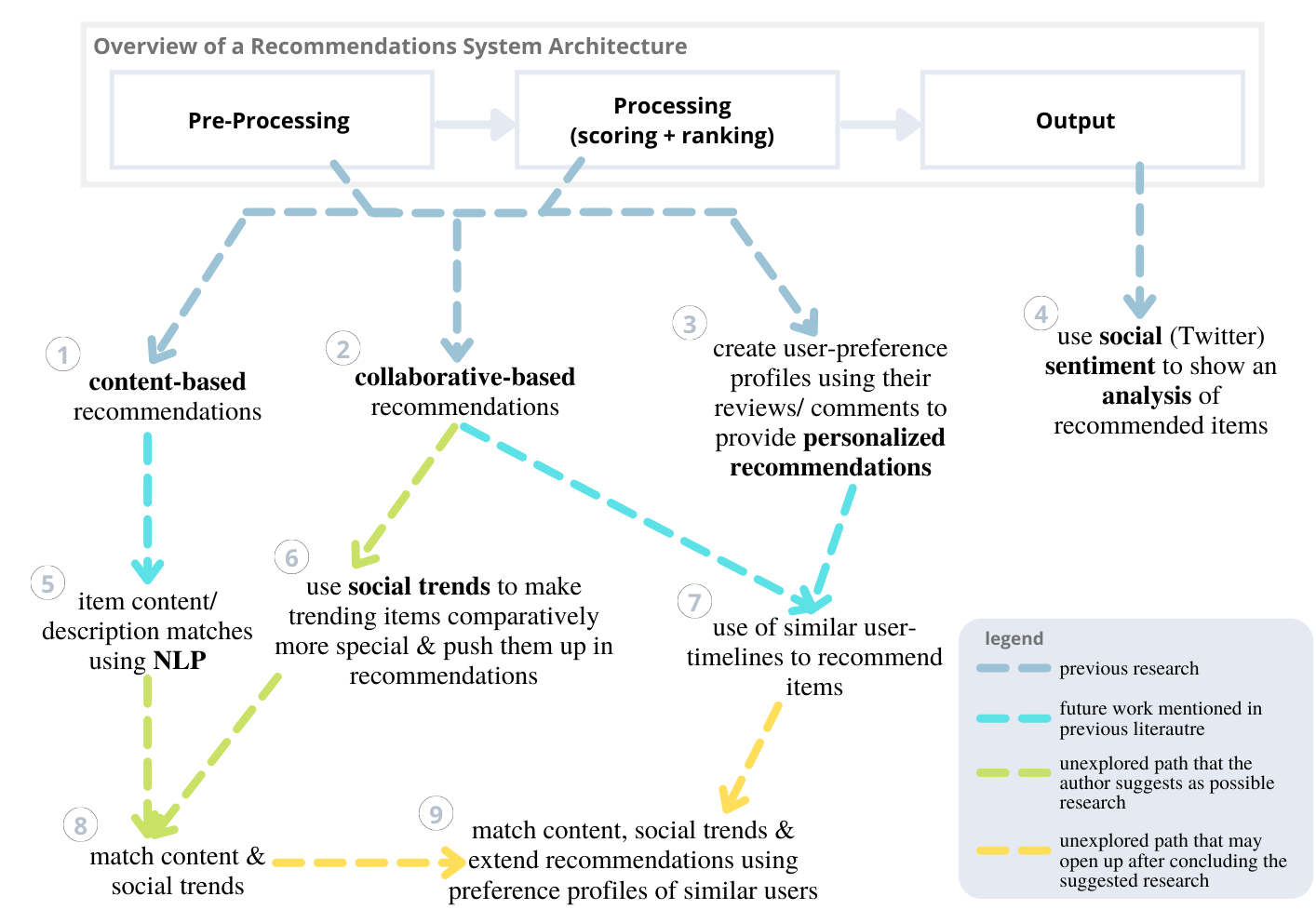}
\caption{Enhancements done to Recommendation Systems using opinion mining techniques \textit{(self-composed)}}
\label{fig:recommendation-opinion-mining-enhancements}
\end{figure}

There have been many attempts to expand the capabilities of Recommendations by making use of public opinion. Collaborative Filtering was one approach to achieve that. Another identified approach was to make use of user data on social media. This has been integrated into Machine Learning-based Hybrid Recommendation Architectures in many ways. In the figure\ref{fig:recommendation-opinion-mining-enhancements}, the author tries to elaborate on the possible technical contribution brought forward in this research.

The figure \ref{fig:recommendation-opinion-mining-enhancements} shows the identified possible points of integration of opinion mining techniques into a Recommendations System.
1, 2 \cite{linden_amazoncom_2003, larry_history_2019}, 3 \cite{cheng_hybrid_2020} \& 4 \cite{ayushi_cross-domain_2018} techniques have been already applied as identified in past literature, while the 7\textsuperscript{th} technique has been mentioned as possible future work from the 3rd technique \cite{chen_user_2019}. Method 5 hasn't been explicitly attempted in recent literature concerning Recommendation Systems, but the data science models used aren't expected to require a lot of tweaking to achieve it, after the feature engineering step is taken care of.

Method 6 has not been identified in previous literature and is expected to align better with the desires circulating market places that don't directly track \& collect user input such as user clicks. This can be extended to method 8. Finally, if methods 7 \& 8 turn out to give promising results, method 9 would be the next step to provide a completely new personalized recommendations architecture that integrates social media trends that are related to the content of the items.

\section{NLP techniques that can be applied to support the integration of opinion mining into Recommendation Systems}

The main NLP techniques that were identified to be useful to be implemented in a system that requires data-mining \& opinion mining techniques were Sentiment Analysis, Named Entity-Recognition, Tokenization, Stemming \& Lammetization; the latter 4 techniques being required for pre-processing scraped data from opinion-mining techniques.

In order to apply these techniques, many past literature (as mentioned in Existing Work), point in the direction of using industrial-grade libraries that utilize \textbf{Recurrent Neural Network (RNN) architectures} such as \textit{SpaCy and NLTK}. The most state-of-the-art models \& techniques that make use of \textbf{Transformer architectures} can be found in the \textit{Hugging Face} library \cite{wolf_transformers_2020}. Transformer models particularly trained to analyze Twitter Sentiment can be identified to be highly accurate \& fast with classifying opinion mining data from social media \cite{noauthor_cardiffnlptwitter-roberta-base-sentiment_nodate, barbieri_tweeteval_2020}.

\section{Common Challenges Faced by Sentiment-aware Recommendations Systems}
One of the most common challenges faced by Recommendation Systems that wish to integrate sentiment analysis into any part of their architecture is the inability to filter out and classify sarcastic comments. This is an area of NLP that needs to be researched further together with the ability to classify the relevant comments.

Extracting all this useful information in real-time in large volumes can be very challenging. This is why several of these systems appear to fail to provide recommendations in real-time.

\section{Practices to be followed to optimize the usage of gathered opinions}
When considering multiple opinions related to a specific topic/ item, they can be combined into one document and processed rather than processing each opinion one by one \cite{nah_opinion_2018}. When doing so, it would be good to have an impact score for each document to make sure that recommendations are biased appropriately towards the opinions of the majority with consideration of the users' opinions.


\section{Evaluation Approaches for Recommendation Systems}

As highlighted in past reviews, evaluating \& benchmarking Recommendation Systems has been a major concern due to the lack of available datasets and questions related to domain-specific approaches/ algorithms used for recommendations.

For the convenience of future work, the following breakdown can be used to evaluate future Recommendations Systems. Especially, those that integrate novel algorithmic, hybrid approaches for recommendations.

\bigbreak
When evaluating Recommendation Systems, we may examine the outcomes produced by the system in two ways.
The first way would be to identify if the system is capable of recommending items that a user may use. The second method would be to identify if the system is capable of recommending items that a user will choose/ use.

The first way to evaluate the outcome can be done by utilizing current data and pre-identified conditions. For the second approach, the evaluation algorithm would require feedback from the public. This can be done by having open beta testing. It would take more time \& effort, but it will be capable of evaluating a system qualitatively on the final goal instead of a possibility.

If we look at evaluating this system from an expected-output performance point of view, \textit{Precision@K (P@K)}, also identified as the \textit{Top-N strategy} in several pieces of literature is the most common method of evaluating a Recommendations System.
This measure and the metrics that have been mentioned below can be used to \textbf{quantitatively} evaluate Recommendation Systems.

\begin{table}[h]
\caption{Evaluation techniques for Recommendation Systems}
\label{tab:evaluation-techniques-table}
\centering
\begin{tabular}{|l|p{0.5\linewidth}|p{0.21\linewidth}|} 
\hline
Measure & Description & Objective Orientation \\ 
\hline
MAE & Measures the average absolute deviation between a predicted rating and the user’s true rating, overall the known ratings. & \multirow{2}{=}{Negatively oriented. Lower, the better.} \\ 
\cline{1-2}
RMSE & A variant of MAE emphasizes large errors by squaring them. &  \\ 
\hline
Precision & The percentage of items in the recommended list that are assessed to be relevant to the user (i.e. it represents the probability that a selected item is relevant). & \multirow{2}{=}{Positively oriented. Higher, the better.} \\ 
\cline{1-2}
Recall & The ratio of relevant items presented by the system to the total number of relevant items available in the items in the system. &  \\
\hline
\end{tabular}
\end{table}

Mean Absolute Error (MAE) \& Root Mean Squared Error (RMSE) is used to measure the accuracy of predicted user ratings (1-5 star ratings) per item, per user. Precision \& recall are used to measure if the system successfully predicts which items the user will select or consume \cite{dayan_recommenders_2011}.

Since the goal of the Recommendations System is to provide the user with multiple options, it is better if the system can produce options across a diverse range. To evaluate the diversity of items across the produced recommendations, \textit{Aggregate diversity} can be measured.

Apart from these metrics, quality-of-service measures such as CPU \& Memory usage can be considered for evaluation as well.

In the review questioning the advancements of Recommendation Systems, \cite{dacrema_are_2019} the author mentions that the lack of used datasets and code bases hinders the ability to properly benchmark and evaluate new research related to Recommendation Systems. The importance of reproducibility of research related to Recommendations Systems has future been elaborated on in the reviews that follow \cite{dacrema_troubling_2021, ferrari_dacrema_critically_2020, dacrema_methodological_2020}.

\section{Conclusion}
In this review, the author has pointed out the reasons to as why baseline Recommendation Systems have been attempted to be improved with the use of ensemble techniques \& hybrid models. Several attempts that past researchers have taken towards implementing such Recommendation models were also discussed and critically evaluated.

Since this review has covered many opinion mining-related work, the NLP techniques that can be applied to these, the common challenges faced and practices that could be followed to optimize the use of gathered opinions for recommendations purposes have been noted. Finally, since evaluating Recommendation Systems has also not been clearly handled in the past, the author has included a summary of several evaluation metrics that can be adopted together with each of their objected orientations.


\printbibliography










\end{document}